\documentclass[conference,11pt]{IEEEtran}
\IEEEoverridecommandlockouts
\usepackage{cite}
\usepackage{amsmath,amssymb,amsfonts}
\usepackage{algorithmic}
\usepackage{graphicx}
\usepackage{textcomp}
\usepackage{xcolor}
\usepackage{float}
\usepackage{url}
\usepackage{pst-node,pst-text,pst-3d,pst-char}
\usepackage{amsmath}

\ifCLASSOPTIONcompsoc
    \usepackage[caption=false, font=normalsize, labelfont=sf, textfont=sf]{subfig}
\else
\usepackage[caption=false, font=footnotesize]{subfig}
\fi

\def\BibTeX{{\rm B\kern-.05em{\sc i\kern-.025em b}\kern-.08em
    T\kern-.1667em\lower.7ex\hbox{E}\kern-.125emX}}
\begin{document}

\title{{\large \vspace*{-3mm} \hspace*{-3mm} 2020 Fifteenth International Conference on Ecological Vehicles and Renewable Energies (EVER)}\vspace{5mm}\\
%%%%%%%%%%% WRITE THE TITLE OF YOUR PAPER HERE AFTER %%%%%%%%%%%
Value of Fleet Vehicle to Grid in Providing Transmission System Operator Services 
\vspace*{-2mm}}

%\title{

\author{\IEEEauthorblockN{Cormac O'Malley \ \ \ \ \      Marko Aunedi   \ \ \ \ \     Fei Teng   \ \ \ \ \     Goran Strbac}
\IEEEauthorblockA{Department of Electrical and Electronic Engineering \\
Imperial College London\\
London, United Kingdom \\
\\}

\vspace{-0.8cm}}

\maketitle
%\IEEEpubidadjcol

\begin{abstract}
%In power systems with high wind generation capacity, frequency response (FR) and inertia requirements often limit the maximum displacement of thermal plants. FR provision from batteries can reduce required thermal generation, increasing the integration of wind power. Commercially owned electric vehicles (EVs) could provide this FR at a lower capital investment than dedicated grid batteries. %move this to intro your abstract is too long 

In this paper a new aggregated model for electric vehicle (EV) fleets is presented that considers their daily and weekly usage patterns. A frequency-constrained stochastic unit commitment model is employed to optimally schedule EV charging and discharging as well as the provision of frequency response (FR) in an electricity system, while respecting the vehicles' energy requirements and driving schedules. Through case studies we demonstrate that an EV with vehicle to grid (V2G) capability can reduce system costs in a future GB electricity grid by up to £12,000 per year, and reduce CO$_2$ emissions by 60 tonnes per year, mainly due to reduced curtailment of wind power. The paper also quantifies the changes in the benefits of fleet V2G resulting from variations in FR delivery time, the penetration of wind or the uptake of alternative flexibility providers. Finally, a battery degradation model dependent on an EV's state of charge is proposed and implemented in the stochastic scheduling problem. It enables significant degradation cost reductions of 16\% with only a 0.4\% reduction of an EV's system value.

%The EVs' response capability is determined by the charger type it is connected to the grid via: vehicle to grid (V2G), smart or dumb. %move to intro 

\end{abstract}


\section*{Nomenclature}
\vspace{-0.1cm}
\subsection*{Indices and Sets}

$n, N$ \hspace{0.43cm}  Index, Set of nodes in the scenario tree. \\
$g, G$ \hspace{0.5cm} Index, Set of generators.\\

\vspace{-0.25cm}
\subsection*{Constants}

$\Delta \tau (n)$ \hspace{0.15cm} Time-step corresponding to node $n$ (h). \\
$\Delta f_{max}$  \hspace{0.12cm} Maximum admissible frequency nadir (Hz). \\
$\pi (n)$ \hspace{0.51cm} Probability of reaching node $n$. \\
$f_0$ \hspace{0.9cm} Nominal frequency of the power grid (Hz). \\
$P_l$ \hspace{0.9cm} Largest power infeed (GW). \\
$RoCoF_{max}$ \hspace{0.1cm} Maximum admissible RoCoF (Hz/s). \\
$T_P$ \hspace{0.8cm} Delivery time of PFR (s).\\
$T_E$ \hspace{0.8cm} Delivery time of EFR (s).\\
$c^{LS}$ \hspace{0.7cm} Value of lost load (£/GW).\\
$T_{ab}(n)$ \hspace{0.25cm} Absolute time of node $n$. \\
$T_{out}$ \hspace{0.61cm} Start time of the workday. \\
$T_{in}$ \hspace{0.8cm} End time of the workday. \\
$C_{out}$ \hspace{0.6cm} Normalised EV SOC at workday start.\\
$C_{in}$ \hspace{0.75cm} Normalised EV SOC at workday end.\\
$N_{EV}$ \hspace{0.55cm} Number of EVs.\\
$c^D$ \hspace{0.9cm} Capacity fade cost (£/\%).\\
$\omega$ \hspace{1.1cm} Capacity fade shoulder SOC value.\\

\vspace{-0.25cm}
\subsection*{Decision Variables}

$P^{LS}(n)$ \hspace{0.13cm} Load shed at node $n$ (GW). \\
$P_d(n)$ \hspace{0.42cm} Discharge rate of EVs at node $n$ (GW).\\
$P_c(n)$ \hspace{0.45cm} Charge rate of EVs at node $n$ (GW).\\
\vspace{-0.25cm}
\subsection*{Linear Expressions of Decision Variables}

$C_g(n)$ \hspace{0.11cm} Operating cost of thermal unit $g$ at node $n$ (£). \\
$E(n)$ \hspace{0.25cm} Normalised SOC of aggregated EVs at node $n$. \\
$Q_{l}(n)$ \hspace{0.15cm} Rate of battery capacity fade at node $n$ (\%/h).\\
$H(n)$\hspace{0.35cm} System inertia post loss of $P_l$ \hspace{-0.2cm} at node $n$ (GWs). \hspace{-1cm} \\
$R_P(n)$ \hspace{0.01cm} Total PFR provision at node $n$ (GW). \\
$R_E(n)$ \hspace{0.01cm} Total EFR provision at node $n$ (GW). \\

\section{Introduction} \label{table1}

Decarbonisation of the UK economy will require large scale integration of renewable energy sources into the electricity sector. The displacement of conventional generation lowers the overall system inertia, increasing the frequency response (FR) needed to maintain the grid frequency within security boundaries \cite{Teng2016}. Currently, because thermal plants themselves provide most FR services and inertia, the extent of their displacement is limited. The plants' inherent minimum stable generation (MSG) could result in significant curtailment of wind power (COWP) in systems with high wind generation capacity\cite{Tuohy2009}. Since the marginal cost of wind power is close to zero, and so are its marginal carbon emissions, this has negative price and emission implications. %citation?

At the same time, a rapid uptake of electric vehicles (EVs) will be required to shift transport’s energy demand away from the combustion of fossil fuels. The long idle times of a typical EV and its inherent storage capability can be exploited to meet the increased need for transmission system operator (TSO) services in low inertia systems \cite{Lund2008}. This has the dual benefit of: reducing system costs  and producing value for the EV owner. Here, TSO services refer to the means by which a grid-connected EV can reduce system costs, such as providing: reserve, fast frequency response and arbitrage. The TSO service capability of an EV is determined by its charger type and the adopted charging regime. In this paper three distinct charging approaches are considered:
\begin{itemize}
    \item Unmanaged, EVs immediately charge to full when plugged in, no TSO services.
    \item Smart, unidirectional power flow from grid to vehicle, demand shifting, FR and reserve via charging reduction.
    \item Vehicle to grid (V2G), bidirectional power flow, arbitrage, improved reserve and FR via fast switching from charge to discharge.
\end{itemize}

Despite widespread recognition that the enhanced FR and demand shifting capabilities of V2G have the potential to reduce system costs much more than through smart charging, the uptake of V2G solutions has been slow. Some of the main barriers include \cite{V2GBritain}: 1)~uncertainty of V2G's value to the system and the sensitivity of that value to system characteristics; 2)~restrictive distribution network constraints that prevent full extraction of domestic V2G’s value; 3)~fears over increased battery degradation; 4)~uncertain market conditions for TSO service providers, which prohibits the formation of a viable V2G business case. %5) V2G chargers are significantly more expensive than smart chargers, by approximately £5 000 according to \cite{V2GBritain}. %check the format for displaying currency values 

The novel contributions of this paper address some of these barriers, listed here:
\begin{itemize}
    \item The capability to model EVs is added to a complex frequency-secured stochastic unit commitment (SUC) scheduling program, allowing detailed quantification of an EV's impact on system cost and emissions.
    \item A novel, linearised battery degradation model based on the empirically derived model in \cite{Naumann2018} is proposed. Costing the degradation explicitly allows to study  how aggregated EV operation might vary if degradation costs are considered explicitly, and the consequence this has on system value.
\end{itemize}
%This paper uses an advanced stochastic unit commitment (SUC) model with frequency security constraints to quantify the system-value of grid connected EVs. This can inform market and policy formation, also justifying the higher charger cost of V2G. Reduced wind shedding, is identified as the main value creation mechanism. The dependence on RES penetration, FR delivery time and existing system flexibility is explored to understand how an EV's value depends on the type of system it is added to.

This paper only considers fleet EVs. A fleet EV belongs to a group of vehicles owned and operated by an organisation. In the UK, 63\% of new road vehicles are bought by fleets \cite{fleet} so they are important in decarbonising road transport. Fleets have predictable usage schedules similar to that of a domestic EV exclusively used for commuting. In addition, fleets are charged centrally, resulting in less constrictive distribution constraints. %seems unlikely that fleet vehicles act like consumer vehicles but I doubt this will be an issue

The structure of this paper is as follows: the UC with frequency security constraints, EV and battery degradation models are described in section \ref{s1}. Section \ref{s2} presents several case studies that illustrate the value of grid-connected EVs. Finally, section \ref{s3} gives the conclusions.

%INSERT A NOVEL CONTRBUTIONS SECTION

\section{Unit Commitment With Frequency Security Constraints}\label{s1}
This section briefly introduces the advanced SUC model used to evaluate the value of different EV configurations to the system. The scheduling model optimises system operation by scheduling operating reserve, Enhanced Frequency Response (EFR), Primary Frequency Response (PFR) and energy production in light of uncertain renewable output. 

\begin{figure}[b]
\centerline{\includegraphics[width=.85\linewidth]{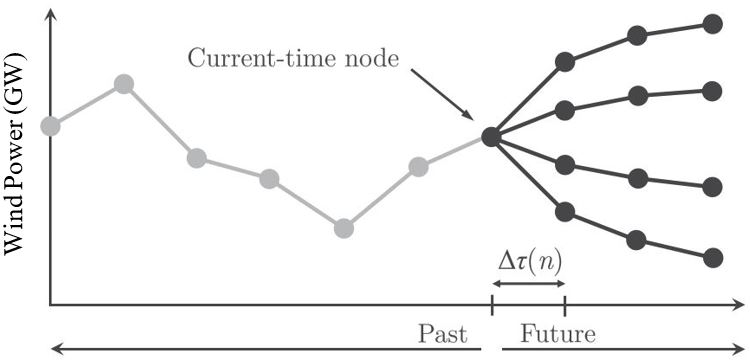}}
\caption{Schematic of a typical scenario tree used in SUC \cite{Badesa2019}.}
\label{tree}

\end{figure}

Reference \cite{Sturt2012} formulates the scheduling problem for a single bus system as a mixed integer linear program that is solved over a multi-stage scenario tree as shown in Fig.~\ref{tree}. The tree nodes are formed from user defined quantiles of the auto-regressive wind model. The probability of reaching a given scenario is used as a weighting in the cost function:
\begin{equation}
\sum_{n\epsilon N} \pi(n) \bigg( \sum_{g\epsilon G} C(n) + \Delta \tau (n) (c^{LS} P^{LS}(n)) \bigg)\label{Cost Function}
\end{equation}

%my knowledge on this stuff is low so Ican't really check it%
Simulations are performed in a similar manner to model predictive control where the complete commitment and dispatch decisions are made every half-hour for a 24h hour period. The simulation implements the optimal decisions at the current node and rolls the system forward by half an hour by updating system states including the actual wind realisation. The process is then repeated. 

Scheduling is subject to the power balance constraint and local-level storage and generator constraints including commitment-time, minimum stable generation and minimum up/down times. Constraints are exhaustively listed in \cite{Sturt2012}.

\subsection{Frequency Security Constraints}\label{AA}

In \cite{Teng2016}, Teng \textit{et al} show that inertia-dependent frequency security constraints can be derived from the swing equation. When these constraints are linearised and applied to the scheduling problem, frequency security following the loss of the largest power infeed is guaranteed at all times. This work was built on in \cite{Badesa} to incorporate FR with different delivery times, optimally allocated by a mixed integer second order cone program. Each FR service is modelled as a linear ramp with a magnitude $R$ GW delivered after a time $T$ s. FR must be delivered for 2 minutes, after which response from slower units, not modelled here, is assumed to bring the frequency back to its nominal value. A simplified version of these constraints that neglect system damping are presented here.

%WHAT IS MISCOP?

\subsubsection{RoCoF}
The largest Rate of Change of Frequency (RoCoF) occurs at the instant of power infeed loss when no FR has been delivered. It is only limited by system inertia:
\begin{equation}
|RoCoF|=\frac{P_l\cdotp f_0}{2H(n)} \leq  RoCoF_{max} \label{RoCoF}
\end{equation}

\subsubsection{Steady State}
For the frequency to stabilise after a loss of generation, the amount of FR available must be at least equal to the power outage:
\begin{equation}
R_E(n) + R_P(n) \geq P_l \label{Steady_State}
\end{equation}

\subsubsection{Nadir}
The frequency Nadir depends on both the available FR and the system inertia: 
\begin{equation}
\bigg(  \underbrace{ \frac{H(n)}{f_0}-\frac{R_E(n) \cdotp T_E}{4 \cdotp \Delta f_{max}} }_{= \ x_1} \bigg) \cdotp \underbrace{ R_P(n) }_{= \ x_2} \geq \frac{(P_l - R_E(n))^2 \cdotp T_P }{4 \cdotp \Delta f_{max}}
\label{Nadir}
\end{equation}
Reference \cite{Badesa} notes that the nonlinear constraint \eqref{Nadir} is in fact a rotated Second Order Cone that is convex because $x_1$ and $x_2$ are non-negative.

\subsection{Modelling of Aggregated EVs}

In our model the aggregated EVs act as one large battery that is disconnected during the workday while the vehicles are in use. The aggregated battery's state of charge (SOC) and power rates are equal to the sum of those for individual EVs. The normalised SOC of an individual EV is therefore the same as the normalised SOC of the aggregate battery. The highly predictable use profile of fleet EVs is fully specified by four parameters: the start/end time of a workday and the SOC at both of these times.

When formulating the optimisation problem the program iterates through the nodes sequentially from 1 to $N$, implementing constraints for each node's specific set of decision variables. The absolute time $T_{ab}(n)$ of node $n$ is known during this phase.

If $T_{ab}(n)$ lies between $T_{out}$ and $T_{in}$ on a weekday, then no charging or discharging is allowed:
\begin{equation}
    P_{d} \leq 0 ,  \ \ \ P_{c} \leq 0
    \vspace{-0.1cm}
\end{equation}
This effectively isolates the battery from the system, preventing any TSO service provision.

In the time interval when the vehicles leave the depot, i.e., when $T_{ab}(n) =T_{out}$, the SOC needs to meet the pre-specified energy requirement for driving: 
\begin{equation}
   E(n) = C_{out}
   \vspace{-0.1cm}
\end{equation}
During time intervals when EVs are being driven, the energy demand for driving reduces the SOC to $C_{in}$ with no injection into the electricity system.
The discharge rate $P_{d}$ of EVs following smart and unmanaged charging is always zero. Additionally, the FR capacity of unmanaged chargers is always constrained to zero and $T_{out}$ is set at a time occurring soon after $T_{in}$, forcing the EVs to charge immediately.
\vspace{-0.05cm}
\subsection{Battery Degradation} \label{degradationsection}
\vspace{-0.05cm}
Battery degradation in EVs manifests itself in two ways: capacity fade ($Q_{L}$), that reduces the maximum SOC; and power fade, which is associated with an increase in the internal resistance (IR) that reduces battery efficiency and limits the system’s power capabilities \cite{Barre2013}. The user experiences an implicit cost of battery degradation through reduced EV performance and the need for more frequent EV replacement. Incorporating a degradation cost into the model allows the net system value of EVs to be calculated more accurately. In addition, explicitly considering the cost of degradation incentivises charging and discharging behaviour with a lower impact on battery life.

Battery degradation phenomena are dependent on multiple interacting factors that make model formation highly complex \cite{Vetter2005}.  Nevertheless, battery ageing can be broadly divided into two components: calendar and cycle ageing. Respectively, the main factors to affect the two types of ageing are: SOC, calendar age and temperature; number of Full Equivalent Cycles (FECs), depth of discharge, charge rate and temperature. %This paper only considers a 10~kW charger attached to 40~kWh EV battery. Cycle ageing is not considered further because at such a low ratio of (dis)charge rate to maximum SOC, calendar ageing dominates cycle ageing \cite{Uddin2017} \cite{Vetter2005}.

This paper only considers a 10~kW charger attached to 40~kWh EV battery. Current mass-produced EV battery cells show cycle stability up to 6,000 FECs before a 20\% capacity loss occurs \cite{battery1} \cite{battery2}. The maximum annual FEC number for any charging regime considered here is 263. Total ageing is a combination of cycle and calendar ageing. Consequently, over the minimum time of approximately 23 years required for the EVs simulated here to reach their maximum cycle number, the contribution of calendar ageing is dominant. For this reason cycle ageing is not considered in the SUC.

From extensive testing of lithium-ion batteries similar to those found in EVs, reference \cite{Naumann2018} proposes a calendar ageing model where $Q_{L}$ and IR only depend on temperature, SOC and age. Here, the temperature and age of individual EVs in the aggregate are unknown and uncontrollable so both of those terms are disregarded. The expression for IR as a function of SOC reported in \cite{Naumann2018} is concave. This too is disregarded due to the added complexity incurred when finding the system’s global optimum, but also because capacity fade is by far the more prevalent metric of EV degradation. The relationship between the capacity fade and the battery SOC is given by the following polynomial expression:
\vspace{-0.1cm}
\begin{equation}
Q_{L}(n) = a_1 (E(n)-0.5)^3 + a_2  \label{Battery Deg}
\vspace{-0.1cm}
\end{equation}

%It is important to emphasise that here, the motivation to include a degradation model is not to give a completely accurate cost per EV. Rather, it is to see how operation might vary when EV degradation is considered and how this ultimately effects an EVs value to the system. Battery SOC is the dominant degradation phenomena controllable at a system level for an aggregate of EVs with low C-rate, so it is the only variable in our degradation model. The cubic coefficients for $Q_L$ depend on temperature and the exact type of battery considered, but the cubic relation generalises to all EV batteries. Here coefficients are chosen that give a reasonable range of degradation rates over the SOC range of interest: 

The coefficients ($a_1$, $a_2$) depend on EV battery age, temperature and type. Exact values of these coefficients are not known for a fleet of EVs so the values chosen here give a reasonable range of degradation rates based on the empirical data reported in \cite{Naumann2018}. It is important to emphasise that the objective of the paper is not to propose a highly accurate model for EV battery degradation cost. Rather, it is to see how operation might vary when the cost of EV battery degradation is explicitly considered and how this might impact the value of EVs to the system.
%\begin{equation}
%Q_{L}(n) = 1.34 \times 10^{-3} (E(n)-0.5)^3 + 4.72 \times 10^{-5}  \label{Battery Deg}
%\end{equation}

\begin{figure}[t]
\vspace{-0cm}
\centerline{\includegraphics[width=\linewidth]{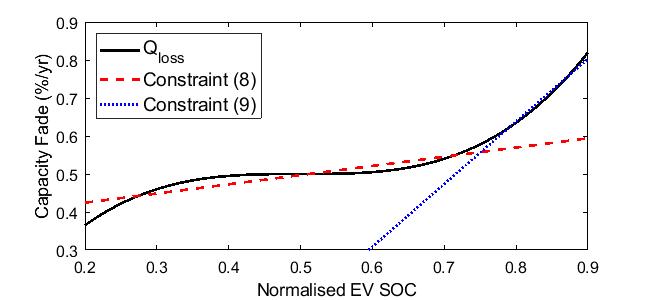}}
\vspace{-0.1cm}
\caption{Cubic relation between capacity fade and SOC, shown with the linear approximation when $\omega=0.76$.\vspace{-3cm}}
\label{degradation}
\vspace{-0.35cm}
\end{figure}

In this paper $a_1=5.70\times10^{-4}$ and $a_2=5.70\times10^{-5}$ are used. This gives absolute annual degradation rates similar to those presented in Fig. 2a of \cite{Naumann2018} for a battery held at SOC = 0.5 and $0^o$C. Furthermore, these coefficients give a ratio of $min(Q_L):max(Q_L)=0.45$ which is within the range presented in Fig. 7a of \cite{Naumann2018}. Fig. \ref{degradation} shows the capacity fade for an individual EV stored at a constant SOC for a year using these coefficients.

$Q_L$ is non-convex so it is approximated by two linear inequality constraints before being added to the SUC:
\vspace{-0.05cm}
\begin{equation}
Q_{L}(n) \geq \alpha_1 E(n) + \alpha_2 \label{constraint 1}
\end{equation}
\begin{equation}
Q_{L}(n) \geq \beta_1 E(n) + \beta_2 \label{constraint 2}
\end{equation}

The coefficients used in this paper are $\alpha_1=2.76\times~10^{-5}$, $\alpha_2=4.29\times10^{-5}$ and $\beta_1=1.88\times10^{-4}$, $\beta_2=-7.76\times10^{-5}$. These are found via linear regression between the boundary SOCs (0.2 and 0.9) and a shoulder value $\omega$. Here $\omega=0.76$ is chosen because the capacity fade at this SOC is halfway between the capacity fade value at the boundary SOCs. Finally, because $Q_{L}$ gives the relative capacity fade of an individual EV it is multiplied by the number of EVs and a cost coefficient $c^D$ in the modified objective function:
%but Fig. \ref{degradation} shows it can be approximated over the SOC range of interest via two linear inequality constraints. Found here via linear regression between the SOC boundaries and $\omega$. When $\omega=0.7$, the constraints are:

%\begin{equation}
%Q_{L}(n) \geq 6.08\times 10^{-5} E(n) + 1.56 \times 10^{-5} \label{constraint 1}
%\end{equation}
%\begin{equation}
%Q_{L}(n) \geq 3.18 \times 10^{-4} E(n) - 1.64 \times 10^{-4} \label{constraint 2}
%\end{equation}
%The feasible set defined by \eqref{constraint 1} and \eqref{constraint 2} is convex and approximates the equality cost because $Q_{L}$ is minimised in the objective function. 

\vspace{-0.2cm}

%\begin{equation}
%\sum_{n\epsilon N} \pi(n) \bigg( \sum_{g\epsilon G} C_g(n) + \delta \tau (n)( c^{LS} P^{LS}(n) +  c^{D}  Q_{L}(n) N_{EV}) \bigg)\label{Cost Function Degradation}
%\end{equation}

\begin{multline}
    \sum_{n\epsilon N} \pi(n) \bigg( \sum_{g\epsilon G} C_g(n) + \delta \tau (n)( c^{LS} P^{LS}(n) \\ +  c^{D}  Q_{L}(n) N_{EV}) \bigg)\label{Cost Function Degradation}
\end{multline}

\vspace{-0.2cm}

\section{Case Studies}\label{s2}

%cut whole swathes of text from here the paper is too long for confernce this is all fine for thesis and maybe even journal but too much for confernce 

To explore the cost reductions offered by different grid-connected EV configurations, the SUC model described in section \ref{s1} was used to run a number of case studies simulating the annual operation of a representative GB 2025 system. A wide range of assumptions has been studied for the number of fleet EV connected to the system, ranging between 50,000 and 300,000. For each fleet EV penetration and each of the three charging regimes the value of EV was found as the difference in annual operating cost between a given fleet EV case and the benchmark case with zero EVs, and then divided with the assumed number of EVs.

The system was assumed to include 55~GW of thermal generation capacity, with the characteristics taken from Table~\ref{table1} of \cite{Badesa2019}. The fleet includes 4 must-run 1.8~GW nuclear plants, thus $P_l$ = 1.8~GW. Two types of FR are considered: EFR with delivery time $T_{E} = 1~s$; and PFR with delivery within $T_P = 10~s$. Batteries (including aggregated EVs) are assumed to provide EFR, while thermal and pumped storage plants provide PFR. In line with GB grid standards the frequency criteria of $RoCoF_{max} =$ 1~Hzs$^{-1}$ and $\Delta f_{max} = $ 0.8~Hz were assumed. A scenario tree that branches at the current node only (refer to \cite{Sturt2012} for further explanation) was used for the SUC with quantiles of 0.005, 0.1, 0.3, 0.5, 0.7, 0.9 and 0.995 considered. Historic electricity demand data profile is used in the model, ranging from 25~GW to 60~GW. Unless otherwise specified there is 30~GW of installed wind capacity and 1~GWh/0.25~GW stationary battery storage on the system. There is also 10~GWh/2.6~GW pumped hydro storage capacity on the system capable of providing 0.5~GW of PFR.

Each EV in the fleet is assumed to have a battery with 40~kWh capacity with a normalised SOC range limited between 0.2-0.9, (dis)charge efficiency of 96\% and maximum (dis)charge of 10~kW. The vehicles are disconnected from the grid between 08:00 and 16:00 during working days, with $C_{out}$ = 0.9 and $C_{in}$ = 0.625. $T_{out}$ for unmanaged chargers was set to 21:00. No EVs other than the fleet vehicles were connected to the system. Only section \ref{degradationstudy} considers battery degradation.

\vspace{-0.05cm}
\subsection{Value of EV for various charging regimes}\label{s2.1}

\begin{figure}[b]
\vspace{-0.4cm}
\centerline{\includegraphics[width=\linewidth]{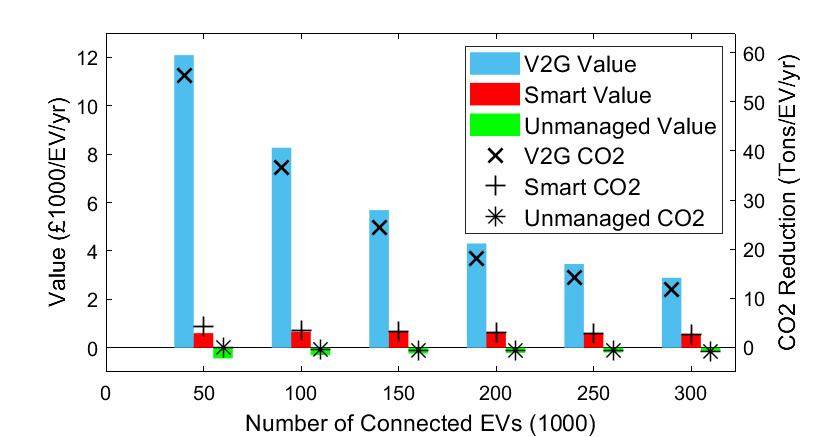}}

\caption{Annual system cost savings and CO$_2$ emission reduction per EV for various charging regimes and EV penetrations.}
\label{CostandCO2}
\end{figure}

 The system values of EVs obtained across various EV uptake levels and charging regimes are shown in Fig.~\ref{CostandCO2}. The results clearly show that V2G is more valuable to the system than Smart for all EV penetrations. With 50,000 EVs on the system each V2G-enabled fleet EV can reduce the system operation cost by £12,000 per annum and CO$_2$ emissions by 60 tonnes per annum whilst meeting all driving requirements. As expected, unmanaged charging increases system costs (i.e., results in a negative value per EV) because it increases the system energy demand without offering any TSO services.

\begin{figure}[t!]
\vspace{-0.3cm}
\centerline{\includegraphics[width=\linewidth]{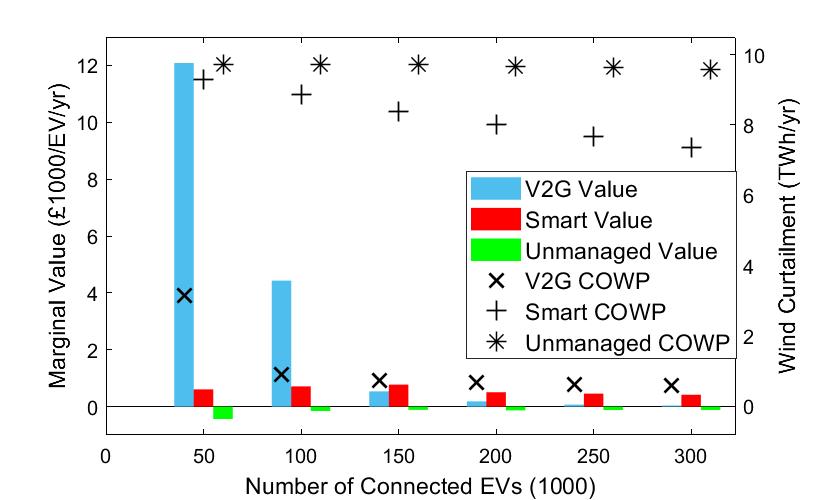}}
\caption{Marginal system value per EV at various EV penetrations plotted with total annual wind curtailment.}
\label{Windshed}
\vspace{-0.4cm}
\end{figure}

Reduced COWP is the main contributor to the observed cost and emission reductions. This is demonstrated in Fig. \ref{Windshed}, which quantifies the marginal system value per EV, obtained by finding the cost difference between two adjacent EV penetrations and dividing it with the difference in EV numbers for the two penetrations. The marginal value of V2G-enabled EVs appears to be broadly proportional to the trend of reducing COWP. 50,000 EVs with V2G enable the system to absorb 6.1~TWh more wind energy than with smart charging, or about 6.7\% of the annual available wind output.

%The marginal cost of wind energy is zero so integrating the most possible is a scheduling priority. Thus, understanding how EV's reduce COWP is vital to understanding their system value and is explained here.

Frequency security constraints effectively impose a limit on the  minimum level of system inertia. In the model presented in this paper inertia is only provided by thermal plants; however, FR can be provided by  both thermal plants and storage. Thermal plants have MSG limits so a minimum system inertia can also be expressed as a requirement for Minimum System Thermal Generation (MSTG). COWP occurs when the sum of  MSTG and the available wind power becomes higher than system demand.

\begin{figure}[b]
\vspace{-0.4cm}
\centerline{\includegraphics[width=\linewidth]{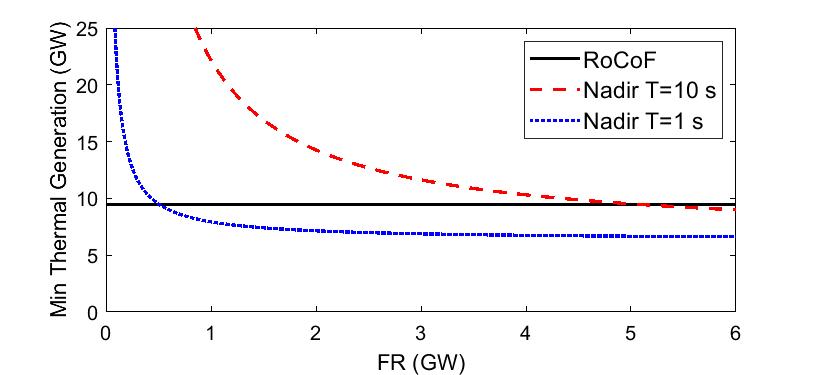}}
\caption{Minimum thermal generation to provide inertia to secure system frequency for a given level of FR. The nadir inertia requirement is derived from \eqref{Nadir} with one FR service (i.e. $R_E=0$) with two different delivery times.}
\label{constraints}
\end{figure}

Fig. \ref{constraints} plots the the relationship between MSTG and available FR capability resulting from constraints \eqref{RoCoF} and \eqref{Nadir}, if there was only one FR service with a delivery time of 1~s or 10~s. The MSTG driven by the RoCoF is independent of FR. Therefore, when sufficient FR is available from EVs to reduce the nadir-driven MSTG below that required to ensure maximum RoCoF, the value of additional EVs reduces significantly because the incremental FR they can provide does not reduce the system's MSTG.

%The nadir inertia limit \eqref{Nadir} is reduced when more FR is available, but the RoCoF inertia limit \eqref{RoCoF} is independent of FR. Fig. \ref{constraints} plots the MSTG due to the nadir constraint if there was only one FR service with either a delivery time of 1s or 10s.
%Wind curtailment occurs when the MSTG and the available wind energy power is higher than system demand. EVs create value in the system by providing EFR to reduce the MSTG required from Nadir. However once the EFR is abundant, around 0.52GW according to Fig. \ref{constraints} the RoCoF MSTG becomes dominant limiting marginal EV value. 

\begin{figure}[t!]
    \centering
  \subfloat[\label{345434}]{%
       \includegraphics[width=1.0\linewidth]{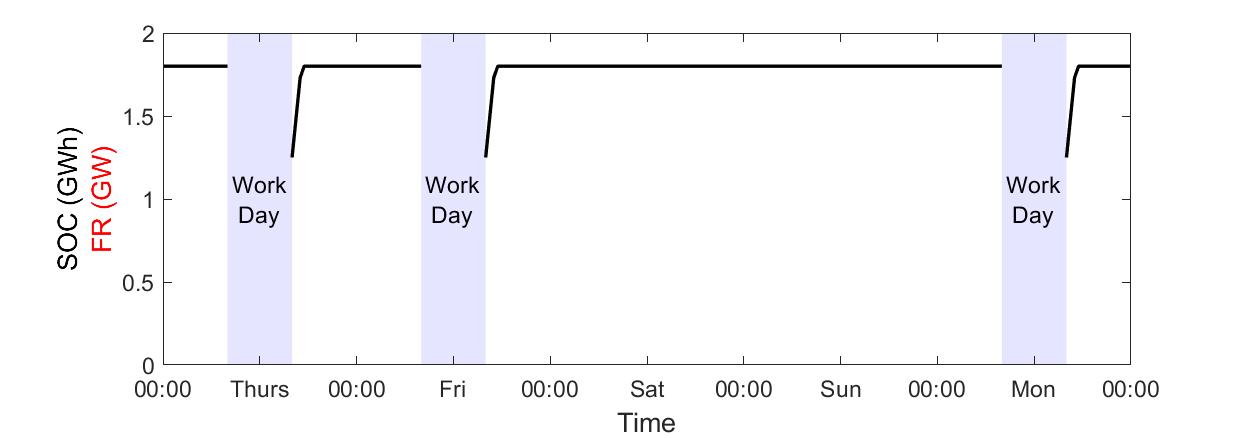}}
          \hfill
 \vspace{-0.7\baselineskip}
    \\
  \subfloat[\label{6557}]{%
        \includegraphics[width=1.0\linewidth]{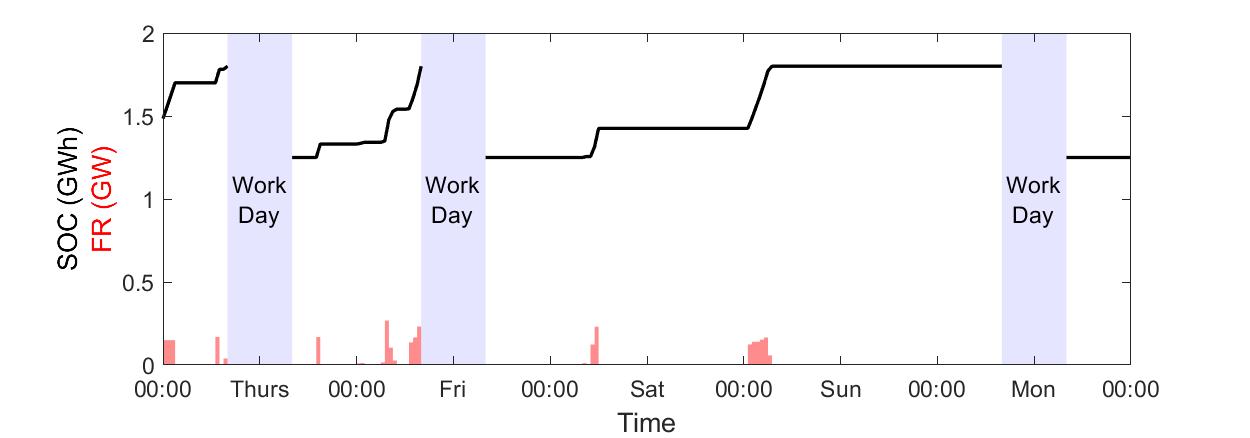}}
    \hfill
     \vspace{-0.7\baselineskip}
    \\
  \subfloat[\label{345434}]{%
       \includegraphics[width=1.0\linewidth]{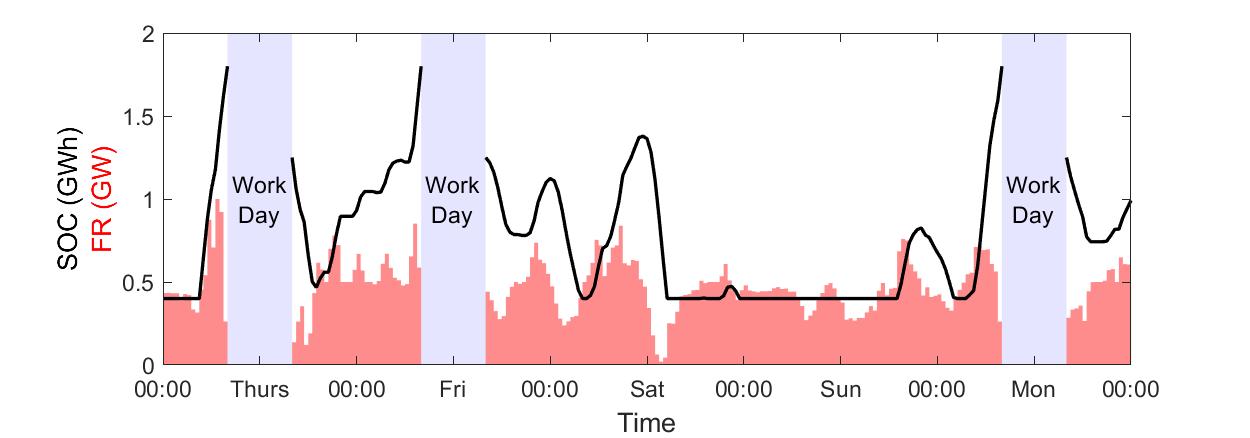}}
          \hfill
    \\
  \caption{SOC and FR provision from EVs over a typical 5-day period for a 50,000 EV fleet with three different charging capabilities: (a) Unmanaged, (b) Smart and (c) V2G.}
  \label{50kthist} 
  \vspace{-0.4cm}
\end{figure}

V2G-connected EVs can provide EFR at all times other than when discharging at maximum rate, while EVs in the Smart charging regime can only provide FR whilst charging, as shown in Fig. \ref{50kthist}. This means that the saturation point in EFR value is reached with far fewer V2G, explaining their higher average value and quickly saturated marginal value. This agrees with the work in \cite{Teng2017} that suggests that at low EV penetrations V2G services are extremely valuable, but at high EV penetrations system FR requirements can largely be met through smart charging.

COWP can also be reduced by charging EVs to increase the system demand during high wind periods, but this is a secondary effect. Although reduced COWP is the dominant value driver, the EVs also create value from: providing reserve thus reducing the need for expensive OCGT plants; increased efficiency of higher loaded online thermal plants and fewer turn on/off events.

%\begin{figure}[htbp]
%\centerline{\includegraphics[width=0.5\linewidth]{V2G300kThist.jpg}}\hfill
%\centerline{\includegraphics[width=0.5\linewidth]{Smart300kThist.jpg}}
%\caption{Notice how not all the FR available is used because of the inertia %requirement}
%\label{300kthist}
%\end{figure}

\subsection{Impact of wind generation and competing flexibility providers on the value of EVs}

System value of EVs is heavily dependent on the characteristics of the system they are added to. This section presents the sensitivity of the EVs' value to renewable penetration, competing flexibility and FR delivery time. The number of EVs in all studies presented in this section was fixed at 50,000. For conciseness only Smart and V2G results are analysed below. 

Table. \ref{table1} shows the change in the EV value as wind capacity increases from 30~GW to 45~GW and 60~GW, as well as the changes in COWP. The value of V2G increases approximately in proportion to the increase in installed wind capacity. The value of Smart remains largely constant over the wind penetrations considered. In line with the discussion in section \ref{s2.1}, reduced COWP is the main value driver. Higher wind capacity increases the amount of time and the magnitude by which the sum of the wind power and the MSTG required for frequency security exceeds demand. This occurs frequently enough with 30~GW of wind capacity that the limited EFR capability of smart chargers is already fully utilised to reduce curtailment, so no additional value is gained. On the other hand, because V2G can provide EFR when idle, MSTG is reduced significantly for most non-workday hours. This means that the instances of COWP are less frequent and of a lower magnitude.
\begin{table}[t]

\caption{Dependence of EV Value on the Penetration of Wind Generation }
\vspace{-0.25cm}
\begin{center}
\begin{tabular}{|c||c|c|c|c|}
\cline{2-5}
\multicolumn{1}{c}{}  &\multicolumn{2}{|c|}{Value (£/EV/yr) }&\multicolumn{2}{|c|}{Wind Curtailment} \\ 
\multicolumn{1}{c}{}  &\multicolumn{2}{|c|}{ }&\multicolumn{2}{|c|}{(TWh/yr)} \\  \hline
Wind Capacity &  \ \ Smart \ \  &  \ \ V2G \ \  &  \ \ Smart \ \ & \ \  V2G \ \ \\ \hline 
30 GW &  608  & 12,088 & \ 9.28 & \ 3.16\\ \hline
45 GW & 590 & 19,769 & 34.92 & 22.19\\ \hline
60 GW & 611 & 23,812 & 69.26 & 53.07\\ \hline
\end{tabular}
\label{renewabletable}
\end{center}
\label{renewabletable}
\vspace{-0.55cm}
\end{table}

\begin{table}[b]
\vspace{-0.5cm}
\caption{Dependence of EV Value on the Penetration of Battery Storage}
\vspace{-0.25cm}
\begin{center}
\begin{tabular}{|c||c|c|} 
\cline{2-3}
 \multicolumn{1}{c}{} &\multicolumn{2}{|c|}{Value (£/EV/yr)} \\
\hline
Battery Capacity (GW:GWh) & \ \ Smart \ \ & \ \ V2G \ \ \\ \hline
0.25:1 & \ 608  & 12,088 \ \\ \hline
0.5:2 & \ 203 & 9,070\\ \hline
1.0:4 & \ \ 47 & 1,921\\ \hline
1.5:6 & -128 & \ \ 475\\ \hline
2.0:8 & -125 & \ \ 335\\ \hline
\end{tabular}
\end{center}
\label{batterytable}
\vspace{-0.2cm}
\end{table}
If EVs are exposed to competition from other flexible sources such as other forms of battery storage, their value can reduce significantly. This is true whether that storage is from other EVs, as in Fig. \ref{CostandCO2}, or from standalone grid-connected batteries, as shown in Table \ref{s1}. This is because when EFR is plentiful, the nadir constraint can be secured easily. After this the MSTG is limited by the thermal plants whose inertia is needed to secure the system RoCoF, upon which additional storage has little effect. It is interesting to note that in Table \ref{s1}, for 1.5~GW and 2~GW of battery storage, the marginal value of EFR is reduced to the point that the cost of the increased energy demand from a smart charger is not offset by its FR provision, resulting in negative system value i.e., positive net system cost per EV.

%I wouldnt bother putting values to the individial pound, it is a model so very unlikely correct to this level. % savings, return on investment and payback time calculations are probably a better option for example 12k saving thanks to system  

Finally, the sensitivity on FR delivery times presented in Fig. \ref{FRtimes} demonstrates that FR provision represents the majority of the system value of V2G-connected EVs. Value per EV is reduced by 95\% from £12,088 to £643 when it provides no FR. This is attributable to the 6.3~TWh increase in annual COWP that occurs when the MSTG is not reduced by the EVs. The speed of FR delivery is also an important value driver: increasing the delivery time from 1~s to 10~s reduces the value of V2G-enabled EVs by about two thirds.

%is mstg stated somewhere?

\vspace{-0.1cm}
\subsection{Battery Degradation}\label{degradationstudy}

\begin{figure}[t]
\centerline{\includegraphics[width=\linewidth]{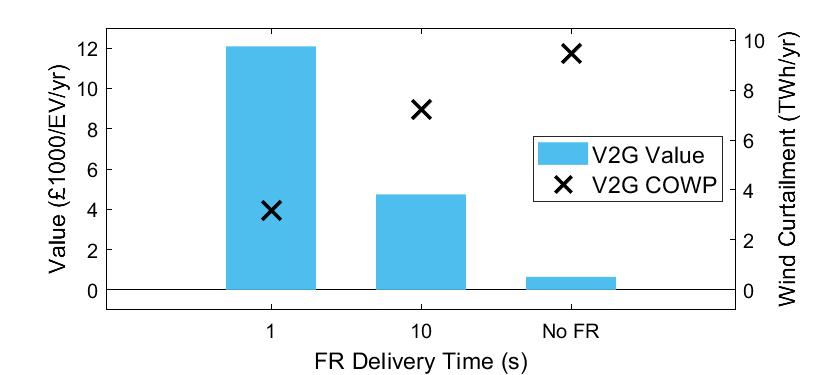}}
\caption{System value of a V2G connected EV with different FR speeds.}
\label{FRtimes}
\vspace{-0.3cm}
\end{figure}

This section applies the battery degradation model developed in section \ref{degradationsection} to a system with 300,000 EVs. A $c^D$ of £1,500/\% was used, implying that a £30,000 EV must be replaced after a 20\% capacity fade. To calculate degradation during the workday driving cycles, the SOC is assumed to be at $C_{out}$ and $C_{in}$ for 4 hours each.

\begin{figure} [b]
\vspace{-0.3cm}
    \centering
  \subfloat[\label{345434}]{%
       \includegraphics[width=1.0\linewidth]{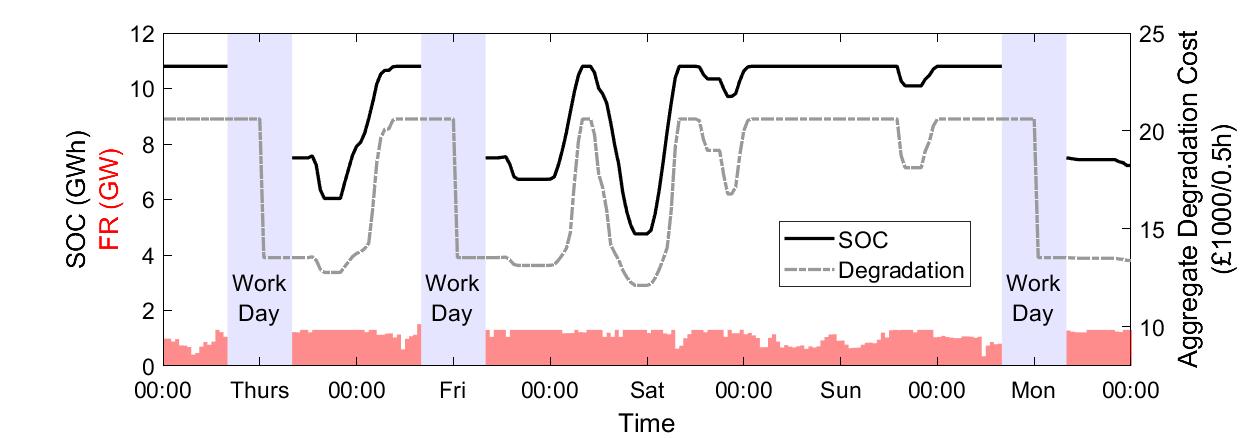}}
          \hfill
 \vspace{-0.7\baselineskip}
    \\
  \subfloat[\label{6557}]{%
        \includegraphics[width=1.0\linewidth]{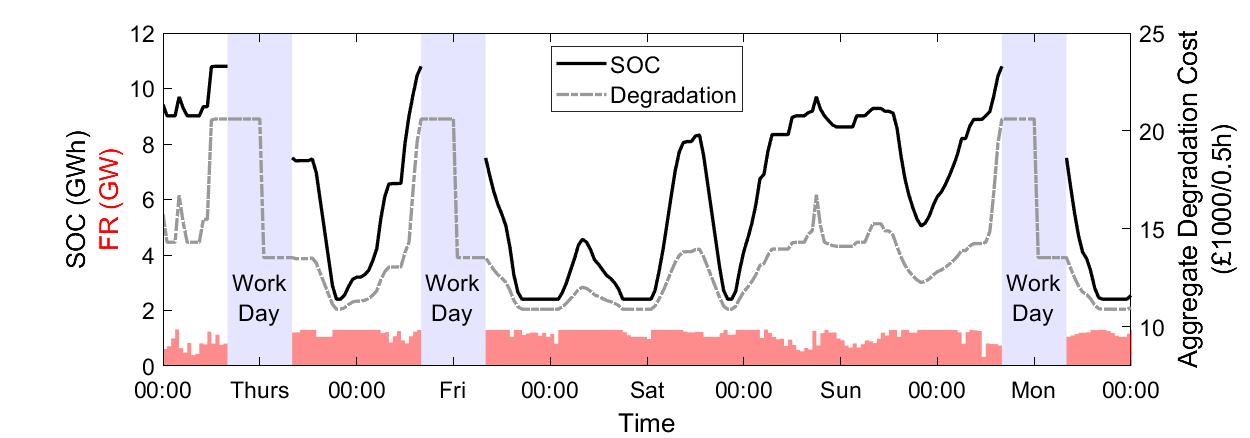}}
    \hfill

  \caption{FR allocation and SOC over a typical 5-day period for 300,000 V2G connected EVs. When the SOC is greater than $\omega$, $Q_{L}$ lies on \eqref{constraint 2}. (a) Degradation is not penalised, cost function \eqref{Cost Function} used. (b) Degradation is penalised, cost function \eqref{Cost Function Degradation} used.}
  \label{300kthist} 
\end{figure}

Fig. \ref{300kthist} shows how V2G-connected EV operation varies over the same 5-day period when degradation is penalised in the cost function. For the case where degradation is penalised, corresponding to the cost function \eqref{Cost Function Degradation}, the EVs' SOC is mostly maintained below 0.75 (9~GWh), as this region corresponds to constraint \eqref{constraint 1}, and therefore results in lower degradation costs. 

\begin{figure}[t!]
\centerline{\includegraphics[width=\linewidth]{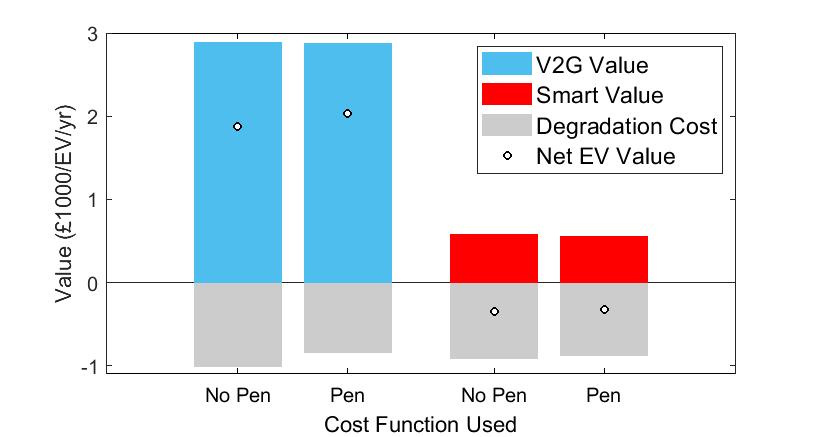}}
\caption{Net value of both smart and V2G connected EVs when degradation is (Pen) and is not (No Pen) penalised in the objective function.}
\label{DegValue}
\vspace{-0.3cm}
\end{figure}

%Inclusion of a degradation cost into the SUC results in significant decreases in battery ageing with limited reduction of an EV's system value. 
As shown in Fig. \ref{FRtimes}, the majority of the system value of EVs derives from the EFR they provide to the system, reducing the COWP as a result. Fig. \ref{300kthist} demonstrates that EFR provision is largely independent of the SOC, so is similar regardless of whether cost function \eqref{Cost Function} or \eqref{Cost Function Degradation} is used. Therefore, incorporating a degradation model into the SUC enables a significant reduction in the EV battery degradation cost of 16.3\%, for only a small decrease in its system value of 0.4\%. This is illustrated in Fig. \ref{DegValue}, which shows the net value of V2G and Smart EVs after factoring in the cost of battery degradation for two cases. The first (No Pen) minimises objective function \eqref{Cost Function} but subsequently adds the degradation cost based on the SOC variations. The second (Pen) is obtained by minimising \eqref{Cost Function Degradation} i.e. explicitly considering the degradation cost. It can be observed that the net value of a V2G-enabled EV increases by £154 if battery degradation is explicitly considered in cost minimisation.

% Fig. \ref{DegValue} shows that operating a V2G connected EV to minimise battery ageing reduces its degradation cost by 16.3\% and only decreases its system value by 0.4\%. This results in an annual net value increase of £154. This limited reduction in system value is because EFR provision is unhindered at a low SOC. 

 %As discussed, battery degradation for aggregated EVs is highly complex and hard to capture in a SUC formulation, but an increased ageing rate at higher SOCs is always true. Hence, these results are valuable as they demonstrate that actions to reduce capacity fade have small negative impacts on system costs relative to the large degradation reductions, thus they are sensible to implement.
 %\cite{Uddin2017}.

\section{Conclusion}\label{s3}

%conclusion should basically just be your abstract again I would try and cut this down

This paper presented a frequency-constrained SUC model that considers aggregated fleet EVs while incorporating the vehicles' driving constraints. The SUC model has been used to demonstrate the impact on system operation cost and carbon emissions when fleet EVs connect to the grid with different charging regimes. FR provision was identified as a key value driver for V2G and Smart charging approaches in a system with high penetration of wind generation, although the EV value was found to be highly sensitive to the uptake of competing flexible options such as standalone batteries and to the speed of FR delivery. A linearised battery degradation model was implemented in the scheduling problem and was used to demonstrate that large decreases in degradation are possible, with only minor reductions in the value of EVs. 

%In this paper an EVs usage schedule was known and completely regular which does not occur in reality. 

Future work will incorporate uncertainty in the driving times and energy usage of EVs into the scheduling process. This will allow value comparison between EVs with different usage profiles, such as fleet or domestic. In addition, work focused on market structures that efficiently convert the demonstrated system cost reductions into revenues for the EV owner will be needed.

%An EV's value to a power system is contingent on it reducing wind curtailment. Hence an EV is most valuable to a system with high renewable penetrations and low storage levels. Reduction of MSTG from thermal plants necessary to provide the inertia to maintain frequency security is the main driver of curtailment reduction. Abundant FR in a system can reduce this MSTG significantly, but is ultimately limited by the RoCoF limit which is only influenced by inertia. This causes a saturation in the value of FR provision from EVs, which is reached earlier with faster FR delivery time. This saturation is reached with only 100 000 V2G chargers, suggesting V2G is most valuable currently, when EV and battery penetrations are low. Once EV penetrations increase the limited FR capability of Smart chargers can meet the systems FR needs.

\section*{Acknowledgment}
The research presented in this paper has been supported by Innovate UK through grant number 104227 (e4Future project) and by National Grid ESO.

\bibliography{V2GConf0302.bib} 
\bibliographystyle{ieeetr}

%\vspace{12pt}

\end{document}